\documentclass[showpacs,preprintnumbers,superscriptaddress,prc,nofootinbib,floatfix,twocolumn]{revtex4-1}
\usepackage[unicode]{hyperref}
\usepackage{amsmath}
\usepackage{amsfonts}
\usepackage{amssymb}
\usepackage{graphicx}
\usepackage{dcolumn}
\usepackage{bm}

\usepackage{color} 

\begin{document}

\title{Richardson-Gaudin integrability in the contraction limit of the quasi-spin}
\author{S. De Baerdemacker}
\email{stijn.debaerdemacker@ugent.be}
\affiliation{Ghent University, Department of Physics and Astronomy, Proeftuinstraat 86, B-9000 Gent, Belgium}
\affiliation{Department of Physics, University of Toronto, Toronto, Ontario M5S 1A7, Canada} %
\affiliation{Department of Physics, University of Notre Dame, Notre Dame, IN 46556-5670, USA}
\date{\today}

\begin{abstract}
\begin{description}
\item[Background] The reduced, level-independent, Bardeen-Cooper-Schrieffer Hamiltonian is exactly diagonalizable by means of a Bethe Ansatz wavefunction, provided the free variables in the Ansatz are the solutions of the set of Richardson-Gaudin equations.  On the one side, the Bethe Ansatz is a simple product state of generalised pair operators.  On the other hand, the Richardson-Gaudin equations are strongly coupled in a non-linear way, making them prone to singularities.  Unfortunately, it is non-trivial to give a clear physical interpretation to the Richardson-Gaudin variables because no physical operator is directly related to the individual variables. 
\item[Purpose] The purpose of this paper is to shed more light on the critical behavior of the Richardson-Gaudin equations, and how this is related to the product wave structure of the Bethe Ansatz.
\item[Method] A pseudo-deformation of the quasi-spin algebra is introduced, leading towards a Heisenberg-Weyl algebra in the contraction limit of the deformation parameter.  This enables an adiabatic connection of the exact Bethe Ansatz eigenstates with pure bosonic multiphonon states.  The physical interpretation of this approach is an adiabatic suppression of the Pauli exclusion principle.  
\item[Results] The method is applied to a so-called ''picket-fence'' model for the BCS Hamiltonian, displaying a typical critical behavior in the Richardson-Gaudin variables.  It was observed that the associated bosonic multiphonon states change collective nature at the critical interaction strengths of the Richardson-Gaudin equations.
\item[Conclusions] The Pauli exclusion principle is the main responsible for the critical behavior of the Richardson-Gaudin equations, which can be suppressed by means of a pseudo deformation of the quasispin algebra.
\end{description}
\end{abstract}

\maketitle

\section{Introduction}
The description of superfluid properties in many-particle systems such as metals \cite{bardeen:1957,vondelft:2001} or atomic nuclei \cite{bohr:1958,dean:2003} involves the process of pairing.  The pairing interaction will cause two particles in time-reversed single-particle states to gain energy by forming a Cooper pair \cite{cooper:1956}.  In the thermodynamic limit, the ground state of the system will constitute a condensate of Cooper pairs, which is well approximated by the Bardeen-Cooper-Schrieffer (BCS) coherent state description of superconductivity \cite{bardeen:1957,rowe:2010}.  While the approximation is tailor made for the thermodynamic limit, it is less appropriate in the mesoscopic finite-size regime \cite{dussel:2007,sandulescu:2008}.  The major source for inaccuracies can be found in the particle-number fluctuations in the BCS state, which become relatively strong with respect to the total number of active particles.  Thus, the ground-state structure of finite-size systems is more intricate than a straightforward BCS condensation of Cooper pairs, so there is a clear call for canonical approaches.  Direct numerical diagonalisation algorithms, such as \emph{e.g.}\ Lanczos, can easily deal with small systems, but they play a loosing game for large finite-size systems, as is the case in \emph{e.g.}\ superconducting metallic nanograins \cite{vondelft:2001} or exotic atomic nuclei \cite{nupecc:2010}, due to the exponential size explosion of the Hilbert space.  

Remarkably, the reduced BCS Hamiltonian, with level-independent scattering terms, was found to be \emph{integrable} \cite{gaudin:1976,cambiaggio:1997}.  A common characteristic of integrable systems is that they support as many (non-trivial) conserved operators, commuting with the Hamiltonian, as there are degrees of freedom in the system \cite{links:2003,dukelsky:2004a}.   Earlier, Richardson had shown that the reduced BCS Hamiltonian can be diagonalised exactly by means of a Bethe Ansatz wavefunction, provided the free variables in the Ansatz (also referred to as rapidities \cite{faribault:2011} or pairons \cite{dukelsky:2004a}) form a solution to the set of non-linear coupled equations, the so-called Richardson-Gaudin (RG) equations \cite{richardson:1963,richardson:1964a}.  Once the RG variables have been determined, all other physical observables can be evaluated from these.

Within the framework of the Algebraic Bethe Ansatz, the eigenstates of an integrable model are given by products of generalised ladder operators \cite{faddeev:1996}.  Whereas these generalised operators can become quite involved in terms of the physical operators acting on the Hilbert space \cite{korepin:1997}, the Bethe Ansatz eigenstates of the reduced BCS Hamiltonian are product states of generalised pair operators.  The factorization of Richardson's Bethe Ansatz into physical operators instigated several investigations regarding its relation to BCS Cooper pairs \cite{pogosov:2011}, projected-BCS Cooper pairs \cite{dussel:2007,sandulescu:2008}, and bosonic excitations \cite{sambataro:2007}.   The clear correspondence between the latter structures and Richardson's exact solution is often obscured by the interpretation of the RG variables.  For one reason, it is well-established that the RG equations display a critical behavior at certain values of the pairing interaction strength, although the energy eigenvalues of the Hamiltonian behave analytically \cite{richardson:1966,rombouts:2004,dominguez:2006}.  Also, no physical observable has been identified yet that can probe the individual RG variables as physical entities, nevertheless, it has been noticed that a qualitative interpretation of the RG variables in terms of the collective behavior of the associated pair creation operators is possible \cite{dussel:2007,sambataro:2007}.

The purpose of the present paper is to shed more light on the critical and collective behavior of the RG variables by making an adiabatic connection with pure-bosonic product state.  This will be done by means of a pseudo-deformation of the $su(2)$ quasispin algebra, related to a suppression of the Pauli exclusion principle.  The integrability of the reduced BCS Hamiltonian and the Richardson-Gaudin formalism will be briefly presented in the next section (Sec.\ \ref{section:rg}).  The following section is divided into four parts, containing a recapitulation of the required ingredients of the $pp$-Tamm Dancoff Approximation (Sec.\ \ref{subsection:tda}), a discussion of the algebraic properties of the pseudo-deformed algebra (Sec.\ \ref{subsection:quasispin}), its embedding into Richardson-Gaudin integrability (Sec.\ \ref{subsection:deformedrg}), and how we can understand more about the critical behavior of the Richardson-Gaudin equations (Sec.\ \ref{subsection:tuningpauli}).  Conclusions are presented in the final section (Sec.\ \ref{section:conclusions}).
\section{Richardson-Gaudin}\label{section:rg}
The reduced BCS Hamiltonian is given by
\begin{equation}\label{rg:hamiltonian}
\hat{H}=\sum_{i=1}^m\varepsilon_i \hat{n}_i + g\sum_{i,j=1}^m\hat{S}^\dag_i \hat{S}_j,
\end{equation}
where the Roman indices $\{i,j=1\dots m\}$ are shorthand for a set of single-particle quantum numbers, with single-particle energy $\varepsilon_i$ and degeneracy $\Omega_i$.  Within the spherical shell model, the degeneracy is $\Omega_i=2j_i+1$ where $j_i$ is the total angular momentum of the level $i$.  When no spherical symmetry is implied, $\Omega_i=2$ refers to the two-fold degeneracy of the intrinsic spin. The Hamiltonian (\ref{rg:hamiltonian}) consists of single-particle operators, counting the number of fermions within a level $i$,
\begin{equation}\label{rg:paircountingoperators}
\hat{n}_i=\sum_{m_i>0}(\hat{a}^\dag_{m_i}\hat{a}_{m_i}+\hat{a}^\dag_{\bar{m}_i}\hat{a}_{\bar{m}_i}),
\end{equation}
and a fermion-pair scattering component, represented by the pair creation/annihilation operators
\begin{equation}\label{rg:paircreationannihilationoperators}
\hat{S}^\dag_i=\sum_{m_i>0}\hat{a}^\dag_{m_i}\hat{a}^\dag_{\bar{m_i}},\quad \hat{S}_i=(\hat{S}^\dag_i)^\dag=\sum_{m_i>0}\hat{a}_{\bar{m}_i}\hat{a}_{m_i},
\end{equation}
with $\hat{a}^\dag_{m_i}$ ($\hat{a}_{m_i}$) the standard fermion creation (annihilation) operators.  The bar notation $\bar{m_i}$ denotes the time-reversed partner of the corresponding operator, \emph{e.g.} $\hat{a}^\dag_{j_i\bar{m}_i}=(-)^{j_i-m_i}\hat{a}^\dag_{j_i,-m_i}$ in the spherical shell model \cite{rowe:2010}.  The set of operators (\ref{rg:paircountingoperators})-(\ref{rg:paircreationannihilationoperators}) span an $su(2)$ quasispin algebra
\begin{equation}\label{rg:quasispin:algebra}
[\hat{S}_i^0,\hat{S}^\dag_j]=\delta_{ij}\hat{S}^\dag_i,\quad[\hat{S}^0_i,\hat{S}_j]=-\delta_{ij}\hat{S}_i,\quad[\hat{S}^\dag_i,\hat{S}_j]=2\delta_{ij}\hat{S}^0_i, 
\end{equation}
where $\hat{S}^0_i=\frac{1}{2}\hat{n}_i-\frac{1}{4}\Omega_i$.  The irreducible representations (irreps) are given by
\begin{equation}
|d_i,\mu_i\rangle=|\tfrac{1}{4}\Omega_i-\tfrac{1}{2}v_i,\tfrac{1}{2}n_i-\tfrac{1}{4}\Omega_i\rangle,
\end{equation}
with $v_i$ the number of unpaired fermions or seniority \cite{talmi:1993}, such that the total number of particles within a level $i$ is given by $n_i=2N_i+v_i$ with $N_i$ the number of pairs.  Richardson's result states that the reduced BCS Hamiltonian (\ref{rg:hamiltonian}) can be diagonalised by means of a product state of generalised pairs acting on the $\bigoplus_{i=1}^m su(2)_i$ lowest weight state $|\theta\rangle=\prod_{i=1}^m|d_i,-d_i\rangle$
\begin{equation}\label{rg:betheansatzstate}
|\psi(\{x\})\rangle=\prod_{\alpha=1}^N\left( \sum_{i=1}^m\frac{\hat{S}^\dag_i}{2\varepsilon_i-x_\alpha}\right)|\theta\rangle,
\end{equation}
provided the set of RG variables $\{x\}$ form the solution of the RG equations
\begin{equation}\label{rg:rgequations}
1+2g\sum_{i=1}^m\frac{d_i}{2\varepsilon_i-x_\alpha}-2g\sum_{\beta=1,\neq\alpha}^N\frac{1}{x_\beta-x_\alpha}=0,
\end{equation}
for $\alpha=1,\dots,N$ with $N$ the total number of pairs.  The curly bracket notation $\{x\}=\{x_1,x_2,\dots,x_N\}$ will be used to denote a set of RG variables, identifying a state.  Once the RG variables $\{x\}$ have been determined, the total energy of the eigenstate is given by
\begin{equation}
E=\sum_{\alpha=1}^N{x_\alpha}+\sum_{i=1}^m \varepsilon_i v_i.
\end{equation}
There are multiple strategies to derive this result, either via Richardson's original approach \cite{richardson:1963,richardson:1964a}, Gaudin's integrability constraints \cite{gaudin:1976}, a linearization of the $R$-matrix in the Algebraic Bethe Ansatz \cite{zhou:2002}, or via a commutator scheme using a Gaudin algebra \cite{ortiz:2005,rowe:2010}.  A brief discussion of the latter approach can be found in Appendix \ref{appendix:derivationrg}, in support of the results of Sec.\ \ref{subsection:deformedrg}.
\section{Hard-core, genuine, and bosons in between}
\subsection{The Tamm-Dancoff Approximation}\label{subsection:tda}
As already mentioned, Richardson's Bethe Ansatz state (\ref{rg:betheansatzstate}) is a simple product state of generalised pair-creation operators, which are highly correlated in a non-linear way by means of the RG equations (\ref{rg:rgequations}).  A sensible approximation would be to relax the intricate correlations described by the RG equations by a simpler and more tractable prescription, retaining the product structure of the wavefunction.  This is exactly what is achieved by certain factorization approximations, such as the projected BCS \cite{ring:2004}, the Random-Phase Approximation \cite{rowe:1970} or the $pp$-Tamm-Dancoff Approximation ($pp$-TDA) (\emph{cfr.}\ Chap.\ 8.2.3 of \cite{ring:2004}). The basic assumption of the $pp$-TDA is that the eigenstates of a Hamiltonian can be considered as harmonic excitations (multiphonon states) of elementary eigen modes (phonons) of the system.  For the pairing Hamiltonian, viable candidates for the elementary excitation modes are given by the 1-pair excitations
\begin{equation}\label{bosonic:tda:eigenvalueequation}
\hat{H}\sum_{i=1}^m Y_{i} \hat{S}^\dag_i|\theta\rangle=\hbar\omega\sum_{i=1}^m Y_{i}\hat{S}^\dag_i|\theta\rangle.
\end{equation}
In the case of the reduced BCS Hamiltonian (\ref{rg:hamiltonian}), the eigenvalue equation (\ref{bosonic:tda:eigenvalueequation}) has a geometric interpretation because the $m$ eigenmodes $\hbar\omega_k$ ($k=1,\dots,m$) can be determined by the zeros of the secular equation of the $pp$-TDA  
\begin{equation}\label{bosonic:tda:secularequation}
1+2g\sum_{i=1}^m\frac{d_i}{2\varepsilon_i-\hbar\omega}=0. 
\end{equation}
The coefficients $Y^k_i$ of the $k$th eigenmode are given by $Y^k_{i}=\frac{1}{2\varepsilon_i-\hbar\omega_k}$, so the $N$-pair wavefunctions can be readily constructed
\begin{equation}\label{bosonic:tda:multiphononstate}
|\psi(\{\hbar\omega\})\rangle=\prod_{\alpha=1}^N\left( \sum_{i=1}^m\frac{\hat{S}^\dag_i}{2\varepsilon_i-\hbar\omega_{k(\alpha)}}\right)|\theta\rangle,
\end{equation}
where $\{\hbar\omega\}$ is shorthand for a set of pair variables $\{\hbar\omega_{k(1)},\dots,\hbar\omega_{k(N)}\}$ with $k(\alpha)$ an integer-valued function selecting the $k$th eigenmode for the $\alpha$th pair.  In the coming, the set $\{\hbar\omega\}$, defining the state (\ref{bosonic:tda:multiphononstate}) will be represented by the TDA distribution $(\nu_1,\nu_2,\dots,\nu_m)$ where $\nu_k$ denotes the number of generalised pairs with eigenmode $\hbar\omega_k$ in the product state (\ref{bosonic:tda:multiphononstate}).  Based on the geometric interpretation of the secular equation, it is seen that the eigenmodes $\hbar\omega_k$ are bound between two subsequent single-particle levels $2\varepsilon_{k-1}<\hbar\omega_k <\varepsilon_{k}$, except for $\hbar\omega_1$, which is bound above by $2\varepsilon_{1}$ but not bound below (see \emph{e.g.}\ Fig.\ 6.6 in \cite{rowe:2010}).  This eigenmode is commonly referred to as the \emph{collective} eigenmode, and a multiphonon state (\ref{bosonic:tda:multiphononstate}) with distribution $(N,0,\dots,0)$ will also be called collective.

\begin{figure}[!htb]
	\includegraphics{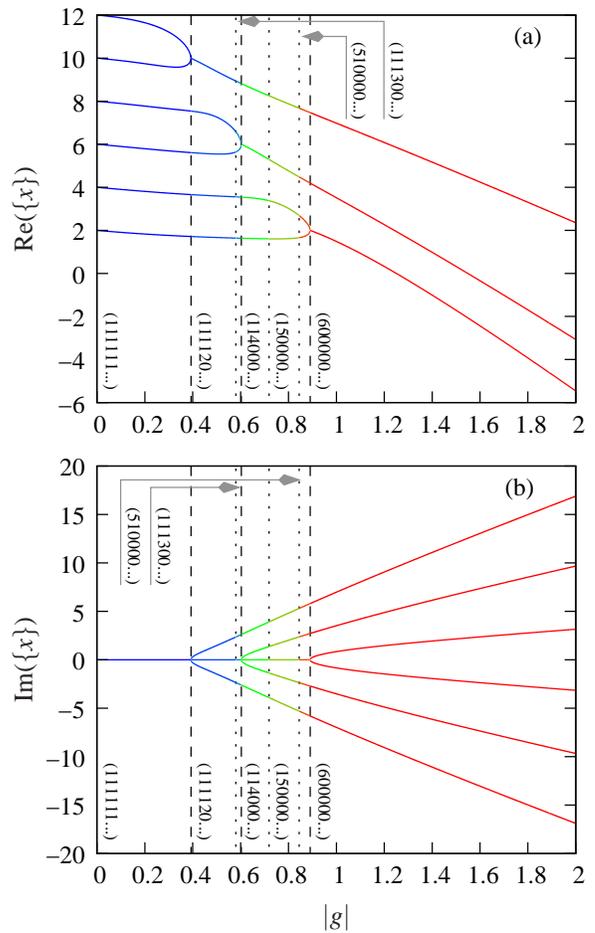}
	\caption{The real (a) and imaginary (b) part of the RG variables of the ''picket-fence'' model, employed in \cite{sambataro:2007}, as a function of the interaction strength $|g|$.  The singular interaction strengths are highlighted by means of dashed lines.  Dotted and dashed lines identify the interaction strengths where the TDA distribution in the contraction limit, denoted by the vectors $(\nu_1,\nu_2,...,\nu_m)$, changes nature (see Sec.\ \ref{subsection:tuningpauli}).}\label{figure:sambataro:rgvariables}
\end{figure}

\begin{figure}[!htb]
	\includegraphics{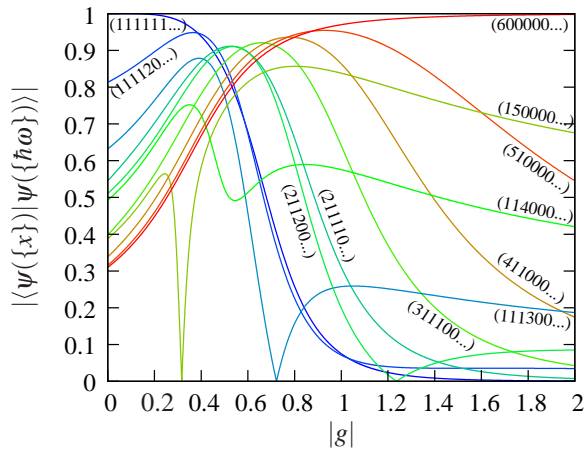}
	\caption{Overlaps of the normalized ground state $|\psi(\{x\})\rangle$ (\ref{rg:betheansatzstate}) of the reduced BCS Hamiltonian (\ref{rg:hamiltonian}) with the normalized product states $|\psi(\{\hbar\omega\})\rangle$(\ref{bosonic:tda:multiphononstate}) for the model described in \cite{sambataro:2007}.  The notation $(i_1,i_2,\dots,i_m)$ denotes how many generalised pairs $i_k$ are chosen with $\hbar\omega_k$.}\label{figure:sambataro:overlaps}
\end{figure}

Given the exact solution of the problem, it is straightforward to check the fitness of the approximation, either by confronting the exact energy spectrum with the expectation value of the Hamiltonian calculated with the multiphonon reference state (\ref{bosonic:tda:multiphononstate}), or by calculating the overlaps of this state with the exact Bethe Ansatz (\ref{rg:betheansatzstate}).  The latter has been done by Sambataro \cite{sambataro:2007}, where a so-called ''picket-fence'' model \cite{hirsch:2002} of 12 uniformly spaced two-fold degenerate single-particle levels has been considered in which 12 paired fermions ($N=6$) are active.  The real and imaginary part of the RG variables are plotted in Fig.\ \ref{figure:sambataro:rgvariables}, showing a typical behavior.  
All RG variables are real valued for sufficiently weak interaction strength ($|g|<0.393$ in Fig.\ \ref{figure:sambataro:rgvariables}).   As the interaction strength increases, the variables recombine two by two to form complex conjugate pairs at the critical interaction strengths $g_c$, until all variables have become complex conjugate pairs ($|g|>0.891$).  The overlaps $\langle\psi(\{x\})|\psi(\{\hbar\omega\})\rangle$ were calculated in \cite{sambataro:2007} for the ground state and a few excited states as a function of the interaction strength $g$.  The conclusion for the ground-state overlaps was that a distinction can be made between 3 different regimes: a weak-coupling ($|g|\lesssim0.25$), intermediate-coupling ($0.25\lesssim |g|\lesssim1.25$) and a strong-coupling regime ($1.25\lesssim |g|$).  It was observed that the overlap with the ground state was largest with the $(1,1,1,1,1,1,0,\dots,0)$ multiphonon state for the weak-coupling regime whereas the overlap was largest with the collective $(6,0,\dots,0)$ multiphonon state in the strong-coupling regime.  For the intermediate regime, the situation was more ambiguous as no single multiphonon state was found to dominate over the others.   To illustrate the situation, the overlaps of the ground state with a selected set of multiphonon distributions are given in Fig.\ \ref{figure:sambataro:overlaps}.
The distributions have been chosen as such that the distribution with maximum overlap with the exact ground state is present in the figure for every interaction strength $g$.  It is readily seen from the figure that several multiphonon states compete for the largest overlap in the intermediate-coupling regime, pointing out that the structure of the exact ground state is quite intricate, opposed to the weak- and strong-coupling regime where a clear picture of the exact ground state emerges, thanks to the dominant overlap of a well-defined multiphonon state.  It can be seen from Fig.\ \ref{figure:sambataro:overlaps} that the maximum overlap changes from  $(1,1,1,1,1,1,\dots)$ to $(1,1,1,1,2,0,\dots)$ at $|g|=0.385$ in the weak-coupling regime and from $(5,1,0,0,0,0,\dots)$ to $(6,0,0,0,0,0,\dots)$ at $|g|=0.872$ in the strong-coupling regime.  These values are remarkably close to respectively the lowest and highest critical points of the RG equations (see Fig.\ \ref{figure:sambataro:rgvariables}), so it would be interesting to know if the critical point in the intermediate regime at $|g|=0.603$ also corresponds to a change in multiphonon nature of the $pp$-TDA.  Unfortunately, the picture is quite blurred in the intermediate regime, so a different approach is needed to enable a clear connection.  This will be discussed in the following subsections.
\subsection{A pseudo deformation of the quasi spin}\label{subsection:quasispin}
Fermionic pairs are similar to bosons in the sense that they mutually commute.  However, opposed to bosons, fermionic pairs feel the presence of other pairs through the Pauli exclusion principle between the constituent fermions.  For these reasons, fermion pairs are sometimes referred to as \emph{hard core} bosons.  It is possible to transform the hard core bosons into genuine bosons by introducing a pseudo deformation of the quasispin algebra.   By redefining the pair creation and annihilation operator such that the influence of the number operator is tunable by means of a deformation parameter $\xi\in[0,1]$
\begin{align}\label{bosonic:quasispin:commutationrelations}
&[\hat{n}_i,\hat{S}^\dag_j(\xi)]=2\delta_{ij}\hat{S}^\dag_j(\xi),\quad[\hat{n}_i,\hat{S}_j(\xi)]=-2\delta_{ij}\hat{S}_j(\xi),\notag\\
&[\hat{S}^\dag_i(\xi),\hat{S}_j(\xi)]=2\delta_{ij}(\xi\tfrac{1}{2}\hat{n}_i-\tfrac{1}{4}\Omega_i),
\end{align}
we arrive at a pseudo deformation of the quasispin algebra, as introduced by Arecchi \emph{et.\ al.}\ \cite{arecchi:1972, gilmore:2008}
\begin{align}\label{bosonic:quasispin:deformedalgebra}
&[\hat{S}^0_i(\xi),\hat{S}^\dag_j(\xi)]=\delta_{ij}\hat{S}^\dag_i(\xi),\quad [\hat{S}^0_i(\xi),\hat{S}_j(\xi)]=-\delta_{ij}\hat{S}_i(\xi)\notag\\ &[\hat{S}^\dag_i(\xi),\hat{S}_j(\xi)]=\delta_{ij}(\xi2\hat{S}^0_i(\xi) + (\xi-1)\tfrac{1}{2}\Omega_i),
\end{align}
with $\hat{S}_i^0(\xi)=\frac{1}{2}\hat{n}_i-\tfrac{1}{4}\Omega_i$ a $\xi$-independent operator and $\hat{n}_i$ retaining its interpretation as number operator.  It is worth emphasizing that no underlying structure is assumed for the creation- and annihilation operators $\hat{S}^\dag_i(\xi)$ and $\hat{S}_i(\xi)$, in contrary to the regular quasispin algebra (see Eq.\ \ref{rg:paircreationannihilationoperators}), so these operators are solely defined by means of the algebra (\ref{bosonic:quasispin:commutationrelations}).   The original quasispin algebra (\ref{rg:quasispin:algebra}) is recaptured in the $\xi=1$ limit of the algebra, and a non-normalized bosonic Heisenberg-Weyl algebra $hw(1)$ is obtained in the $\xi=0$ contraction limit \cite{arecchi:1972}.  It is preferable to employ the term \emph{pseudo} deformation as the deformed algebra (\ref{bosonic:quasispin:deformedalgebra}) effectively spans an $su(2)_\xi$ algebra, defined by the generators
\begin{align}\label{bosonic:quasispin:su2again}
&\hat{A}^\dag_i(\xi)=\tfrac{1}{\sqrt{\xi}}\hat{S}^\dag_i(\xi),\quad  \hat{A}_i(\xi)=\tfrac{1}{\sqrt{\xi}}\hat{S}_i(\xi)\\
&\hat{A}^0_i(\xi)=\hat{S}^0_i(\xi)+(1-\tfrac{1}{\xi})\tfrac{1}{4}\Omega_i=\tfrac{1}{2}\hat{n}_i-\tfrac{1}{4\xi}\Omega_i.
\end{align}
Thanks to the explicit form of the deformed Cartan operator $\hat{A}^0_i(\xi)$, it is possible to assign a loose physical interpretation to the deformation parameter $\xi$.  The action of the Cartan operator on the lowest weight pair vacuum state $|d_i(\xi),-d_i(\xi)\rangle$ yields the following relation for the $su(2)_\xi$ irrep label 
\begin{equation}\label{bosonic:quasispin:deformedirreplabel}
d_i(\xi)=\tfrac{1}{4\xi}\Omega_i-\tfrac{1}{2}v_i.
\end{equation}
As a result, the deformation parameter $\xi$ opens up a given shell by increasing the effective degeneracy $\Omega_i/\xi$, admitting more pairs than strictly allowed in the original irrep $d_i(1)=\tfrac{1}{4}\Omega_i-\tfrac{1}{2}v_i$.  In the contraction limit of $\xi\rightarrow0$, the effective degeneracy reaches infinity, which is consistent with a bosonic interpretation.  It should be emphasized that the assignment of $d_i(\xi)$ as an $su(2)_\xi$ irrep label is only physically meaningful as long as the deformation parameter is $\xi=\frac{1}{1+2k/\Omega_i}$ with $k\in\mathbb{N}$; otherwise, the irreps are no longer unitary \cite{wybourne:1974}.  Nevertheless, this requirement can be relaxed in the framework of Richardson integrability, as everything is well-defined from the commutation relations (\ref{bosonic:quasispin:commutationrelations}) and the action of the operators on the pair vacuum for the deformation parameter over the regime $\xi\in[0,1]$.  
\begin{figure}[!htb]
	\includegraphics{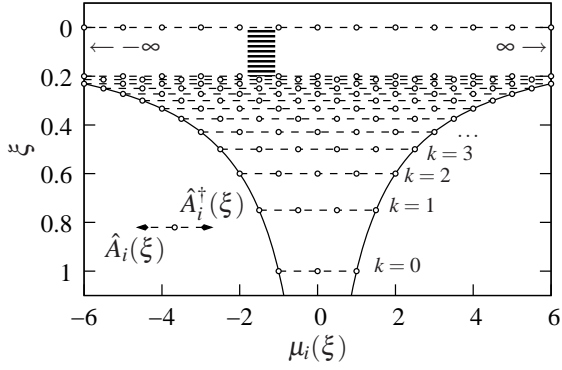}
	\caption{An illustration of the $su(2)_\xi$ quasispin irreps with label $d_i(\xi)$ for $\Omega_i=6$ and $v_i=1$.  The unitary irreps, at $\xi=\frac{1}{1+2k/\Omega_i}$ with $k\in\mathbb{N}$, are denoted by open circles, connected by dashed lines.  The lines connecting the lowest and highest weights with $\mu_i(\xi)=\mp d_i(\xi)$ are drawn in full.  For $\xi=1$, the $d_i(1)=1$ irrep is retained.  The operators $\hat{A}^\dag_i$ and $\hat{A}_i$, depicted in the $su(2)_\xi$ root diagram, are defined in Eq.\ (\ref{bosonic:quasispin:su2again}) }\label{figure:irreplabels}
\end{figure}
Fig.\ \ref{figure:irreplabels} illustrates how the $su(2)_\xi$ irrep label $d_i(\xi)$ grows with decreasing $\xi$, until the contraction limit, where the algebraic structure changes from $su(2)_\xi$ to $hw(1)$.

The algebra (\ref{bosonic:quasispin:su2again}) becomes ill-defined in the contraction limit $\xi=0$.  For this limit, it is better to resort to the original algebra.  By defining the normalized operators
\begin{equation}\label{bosonic:quasispin:heisenbergweyl}
\hat{s}^\dag_i=\sqrt{\tfrac{2}{\Omega_i}}\hat{S}^\dag_i(0),\quad\hat{s}_i=\sqrt{\tfrac{2}{\Omega_i}}\hat{S}_i(0),\quad \hat{s}^0_i=\hat{S}^0_i(0),
\end{equation}
the bosonic nature of the creation/annihilation operators in the contraction limit becomes apparent
\begin{equation}
[\hat{s}^0_i,\hat{s}^\dag_j]=\delta_{ij}\hat{s}^\dag_i,\quad [\hat{s}^0_i,\hat{s}_j]=-\delta_{ij}\hat{s}_i,\quad[\hat{s}_i,\hat{s}^\dag_j]=\delta_{ij}.
\end{equation}
The operators $\hat{s}^0_i$ can be considered equivalent to the number operators $\hat{s}^0_i\equiv \hat{s}^\dag_i\hat{s}_i+\frac{1}{2}v_i-\frac{1}{4}\Omega_i$.
In the following subsection, the integrability of the reduced BCS Hamiltonian (\ref{rg:hamiltonian}) will be investigated using the present pseudo deformation of the quasispin algebra.
\subsection{The pseudo deformed RG equations}\label{subsection:deformedrg}
The key observation is that the reduced BCS Hamiltonian (\ref{rg:hamiltonian}) remains integrable when the deformed commutation relations are used.  The Hamiltonian 
\begin{equation}\label{bosonic:deformedrg:hamiltonian}
\hat{H}(\xi)=\sum_{i=1}^m\varepsilon_i \hat{n}_i + g\sum_{i,j=1}^m\hat{S}^\dag_i(\xi) \hat{S}_j(\xi), 
\end{equation}
is diagonalizable by means of the product state
\begin{equation}\label{bosonic:deformedrg:betheansatzstate}
|\psi(\{x(\xi)\})\rangle=\prod_{\alpha=1}^N\left( \sum_{i=1}^m\frac{\hat{S}^\dag_i(\xi)}{2\varepsilon_i-x_\alpha(\xi)}\right)|\theta\rangle,
\end{equation}
provided the set of deformed RG variables $\{x(\xi)\}$ form the solution of the set of deformed RG equations
\begin{equation}\label{bosonic:deformedrg:rgequations}
1+2g\xi\sum_{i=1}^m\frac{d_i(\xi)}{2\varepsilon_i-x_\alpha}-2g\xi\sum_{\beta=1,\neq\alpha}^N\frac{1}{x_\beta-x_\alpha}=0.
\end{equation}
Remark that the value $\xi d_i(\xi)=\tfrac{1}{4}\Omega_i-\xi\tfrac{1}{2}v_i$ is well-defined and finite over the whole regime $\xi\in[0,1]$.  There are two ways to obtain this result.  The first method is rather straightforward, as discussed in Appendix \ref{appendix:derivationrg}.  One simply commutes the Hamiltonian (\ref{bosonic:deformedrg:hamiltonian}) through the product state (\ref{bosonic:deformedrg:betheansatzstate}) using the commutation relations (\ref{bosonic:quasispin:commutationrelations}) (or equivalently (\ref{bosonic:quasispin:deformedalgebra})), and requires the non-diagonal byproducts to vanish.  The second approach employs the effective $su(2)_\xi$ algebra (\ref{bosonic:quasispin:su2again}) for $\xi\in]0,1]$.  The Hamiltonian (\ref{bosonic:deformedrg:hamiltonian}) can be rewritten as
\begin{equation}
\hat{H}(\xi)=\sum_{i=1}^m\varepsilon_i(2\hat{A}^0_i(\xi)+\tfrac{1}{4\xi}\Omega) + g\xi\sum_{i,j=1}^m\hat{A}^\dag_i(\xi) \hat{A}_j(\xi),
\end{equation}
which can be interpreted as a reduced BCS Hamiltonian, entirely expressed in terms of $su(2)_\xi$ generators, with a scaled coupling constant $g\xi$ and deformed $su(2)_\xi$ irrep labels $d_i(\xi)$ (\ref{bosonic:quasispin:deformedirreplabel}).  Filling out these values in the regular RG equations (\ref{rg:rgequations}) immediately yields the deformed RG equations (\ref{bosonic:deformedrg:rgequations}). The irrep labels $d_i(\xi)$ are not well-defined for $\xi=0$, enforcing a separate assessment of the contraction limit.  The reduced BCS Hamiltonian in this limit is expressible in terms of $hw(1)$ generators (\ref{bosonic:quasispin:heisenbergweyl})
\begin{equation}
\hat{H}(0)=\sum_{i=1}^m\varepsilon_i(2\hat{s}^0_i+\tfrac{1}{2}\Omega_i)  + g\sum_{i,j=1}^m\tfrac{1}{2}\sqrt{\Omega_i\Omega_j}\hat{s}^\dag_i \hat{s}_j.
\end{equation}
The correspondence between $\hat{s}^0_i$ and the bosonic number operator makes the Hamiltonian essentially a one-body bosonic operator
\begin{equation}
\hat{H}(0)=\sum_{i,j=1}^m\left(2\varepsilon_i\delta_{ij} + \tfrac{1}{2}g\sqrt{\Omega_i\Omega_j}\right)\hat{s}^\dag_i \hat{s}_j + \sum_{i=1}^m\varepsilon_i v_i. 
\end{equation}
This Hamiltonian can be brought in diagonal form by means of a simple unitary transformation of the boson operators $\hat{b}^\dag_i=\sum_{k=1}^mU_{ik}\hat{s}^\dag_k$
\begin{equation}\label{bosonic:deformedrg:bosonichamiltonian}
\hat{H}(0)=\sum_{i=1}^m\hbar\omega_i\hat{b}^\dag_i \hat{b}_i + \sum_{i=1}^m\varepsilon_i v_i, 
\end{equation}
with the eigenmodes $\hbar\omega_i$ being the roots of the secular equation
\begin{equation}\label{bosonic:deformedrg:rgequationscontractionlimit}
 1+\frac{g}{2}\sum_{i=1}^m\frac{\Omega_i}{2\varepsilon_i-\hbar\omega}=0.
\end{equation}
This result agrees with the $\xi\rightarrow0$ contraction limit of the deformed RG equations (\ref{bosonic:deformedrg:rgequations}) as $\lim_{\xi\rightarrow0}\xi d_i(\xi)=\frac{1}{4}\Omega_i$, pointing out that the $su(2)_\xi$ part ($\xi\in]0,1]$) and the $hw(1)$ limit ($\xi=0$) are consistent within the pseudo deformation formalism.  A $\xi=0$ multiphonon state can be represented in multiple ways
\begin{align}\label{bosonic:deformedrg:betheansatzstatecontractionlimit}
|\psi(\{x(0)\})\rangle&=\prod_{\alpha=1}^N\hat{b}^\dag_{k(\alpha)}|\theta\rangle=\prod_{k=1}^m(\hat{b}^\dag_k)^{\nu_k}|\theta\rangle\\
&=\prod_{k=1}^m\left(\sum_{i=1}^m\frac{\sqrt{\Omega_i}\hat{s}^\dag_i}{2\varepsilon_i-\hbar\omega_{k}}\right)^{\nu_k}|\theta\rangle,
\end{align}
with $k(\alpha)$ the integer-valued function defined as before, selecting which eigenmode $k$ with energy $\hbar\omega_k$ is assigned to the $\alpha$th phonon.  The total number of phonons with eigenmode $\hbar\omega_k$ in the multiphonon state is given by $\nu_k$, so we can represent a multiphonon state (\ref{bosonic:deformedrg:betheansatzstatecontractionlimit}) by means of a vector $(\nu_1,\nu_2,\dots,\nu_m)$.  This definition is consistent with the definition given in Sec.\ \ref{subsection:tda}, with the only difference that the eigenmodes are here genuine bosons instead of generalised pairs.

In conclusion, the RG equations (\ref{rg:rgequations}) of the reduced BCS Hamiltonian (\ref{rg:hamiltonian}) can be continuously transformed, by means of a pseudo deformation of the quasispin algebra (\ref{bosonic:quasispin:deformedalgebra}), into the secular equation (\ref{bosonic:deformedrg:rgequationscontractionlimit}) of a bosonic one-body Hamiltonian (\ref{bosonic:deformedrg:bosonichamiltonian}) bearing a strong resemblance with the secular equation (\ref{bosonic:tda:secularequation}) of the $pp$-TDA.  The correspondence between the secular equations (\ref{bosonic:tda:secularequation}) and (\ref{bosonic:deformedrg:rgequationscontractionlimit}) can be understood from a physical point of view.  As the eigenmodes of the $pp$-TDA are one-pair excitations, they behave essentially bosonic, so they will be structurally equivalent to the eigenmodes of the bosonic Hamiltonian (\ref{bosonic:deformedrg:bosonichamiltonian}).   The only difference between (\ref{bosonic:tda:secularequation}) and (\ref{bosonic:deformedrg:rgequationscontractionlimit}) is the presence of the seniority $v_i$ in the $pp$-TDA.  This is a direct consequence of how the Pauli exclusion principle is applied in both formalisms.  Throughout the construction, the operator $\hat{n}_i$ was required to retain its interpretation as a number operator.  Therefore, the number of particles occupying the pair vacuum $|\theta\rangle$ equals the number of unpaired particles ($\sum_{i=1}^mv_i$), independent from the deformation parameter.  The regular fermion pairs will feel the presence of unpaired particles, occupying part of the free Hilbert space for the fermion pairs, in contrast to the pure bosons, who are insensitive to the presence of any other particles. 

The advantage of the present construction with the deformed RG equations is that one can make a one-to-one connection between the exact eigenstates of the reduced BCS Hamiltonian and the multiphonon states by adiabatically tuning the Pauli exclusion principle using the deformation parameter $\xi$.  This will be discussed next by means of the picket-fence model introduced in the previous subsection.
\subsection{Tuning the Pauli principle}\label{subsection:tuningpauli}
The deformation parameter $\xi$ of the pseudo deformed quasispin algebra (\ref{bosonic:quasispin:deformedalgebra}) can be used to tune adiabatically the effect of the Pauli exclusion principle.  So, it is now possible to gradually turn off the Pauli principle in a given exact eigenstate, and observe unambiguously to which multiphonon state it will evolve.  The other way around, the exact eigenstates can be reconstructed from physically relevant multiphonon states by reintroducing the exclusion principle.

Before applying these ideas to the picket-fence model of Sec.\ \ref{subsection:tda}, it is instructive to study a few limiting cases.  Much of our understanding of the general features of the RG equations has been derived from these limiting cases, such as \emph{e.g.}\ the weak-coupling  \cite{ortiz:2005} and strong-coupling regime \cite{yuzbashyan:2007}, or the thermodynamic limit \cite{richardson:1977,roman:2002}.  The RG variables in the weak- and strong-coupling regime are related to the roots of the associated Laguerre polynomials using the elegant Heine-Stieltjes (HS) correspondence \cite{stieltjes:1885,szego:1975}.  In general, the HS correspondence relates polynomial solutions of a type of 2nd-order ordinary differential equations with the roots of a set of non-linear coupled algebraic equations.  The RG equations can be nicely embedded within the HS formalism, which has lead to recent developments in the numerics of the RG equations \cite{faribault:2011,pan:2011,elaraby:2012,marquette:2012}.  For some selected cases, the correspondence leads to simple approximate expressions for the RG variables in terms of the roots of special functions, as will be discussed next.  Details can be found in Appendix \ref{appendix:limitsdeformedrg}.

The Heisenberg-Weyl algebraic structure of the contraction limit $\xi=0$ is quite different from the effective quasispin structure for $\xi\in]0,1]$.  Therefore, it would be interesting to see how the RG variables change as the Pauli exclusion is turned on infinitesimally little ($0<\xi\ll1$).  Let  $(\nu_1,\nu_2,\dots,\nu_m)$ be the distribution of a multiphonon state (\ref{bosonic:deformedrg:betheansatzstatecontractionlimit}) in the contraction limit.  In this limit we start from $x_\alpha(0)=\hbar\omega_{k(\alpha)}$ for the RG variables, with $k(\alpha)$ the integer-valued function as defined before.  Using the HS correspondence, it is shown in Appendix \ref{appendix:limitsdeformedrg:nearcontraction} that infinitesimally cranking up the Pauli principle ($\xi\ll1$) affects the RG variables as follows
\begin{equation}\label{bosonic:deformedrg:rgvariablesnearcontractionlimit}
x_\alpha(\xi)=\hbar\omega_{k(\alpha)}+i\sqrt{\tfrac{2\xi}{a_{k(\alpha)}}}z_{\nu_{k(\alpha)},l(\alpha)}+\mathcal{O}(\xi),
\end{equation}
with $z_{\nu,l}$ the $l$th root of the Hermite polynomial $H_\nu(z)$, and $a_k=\frac{1}{2}\sum_{i}\frac{\Omega_i}{(2\varepsilon_i-\hbar\omega_k)^2}$ a $\xi$-independent constant. The integer-valued function $l(\alpha)$ picks a Hermite root for each RG variable, with the only condition that all roots must be distinct.  In other words, the degeneracy in the $\nu_k$ phonons with eigenmode $\hbar\omega_k$ is immediately lifted by turning on the deformation parameter $\xi$.  The dispersion in the RG variables scales with $\sqrt{\xi}$, pointing out that the transition from $hw(1)$ into $su(2)_\xi$, associated with the introduction of the Pauli principle, is both analytic and non-perturbative.  Also, $a_k$ is a positive constant, and the roots of the Hermite polynomials are all real-valued, showing that all (but one) RG variables immediately become complex valued for $\nu_k$ even (odd).  Therefore, we can conclude that it is in fact the Pauli principle pushing those pairs associated with the same eigenmode into the complex plane.  It is worth mentioning that a HS correspondence with the Hermite polynomials has been identified earlier in the weak-coupling limit of the 1D $\delta$-interacting Lieb-Liniger bose gas \cite{gaudin:1971}.  Moreover, a connection with the strong-coupling regime of the reduced BCS Hamiltonian has been established \cite{batchelor:2004}, however there appears to be no connection with the pseudo deformation scheme of the present article.

\begin{figure}[!htb]
 \includegraphics{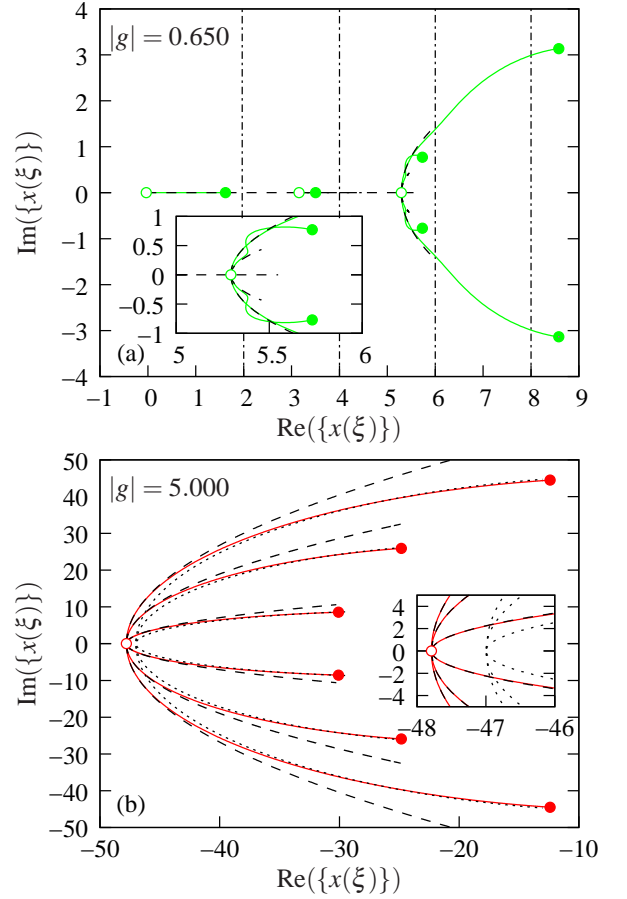}
 \caption{The path of the deformed RG variables $\{x(\xi)\}$ in the complex plain for the  picket-fence model of Sec.\ \ref{subsection:tda} with $|g|=0.650$ (a), and $|g|=5.000$ (b).  the path, given by full lines for each variable, starts from the exact RG variables $\{x(1)\}$ (filled circles), and ends at to the bosonic eigenmodes $\{x(0)\}=\{\hbar\omega\}$ (open circles).  The near-contraction approximation is given by dashed lines, and the strong-coupling limit approximation is given in dotted lines (only for (b)).  The inset in both figure shows a close-up of the path around the near-contraction limit.  It can be remarked that the near-contraction limit predictions and the numerically obtained deformed RG variables nearly coincide at the near-contraction limit for $|g|=5.000$.  The single-particle levels at $2\varepsilon_i$ ($i=1..m$) are also included in dashed-dotted lines as reference points for the multiphonon states. }\label{figure:laguerrehermite}
\end{figure}

\begin{figure*}[!htb]
 \includegraphics{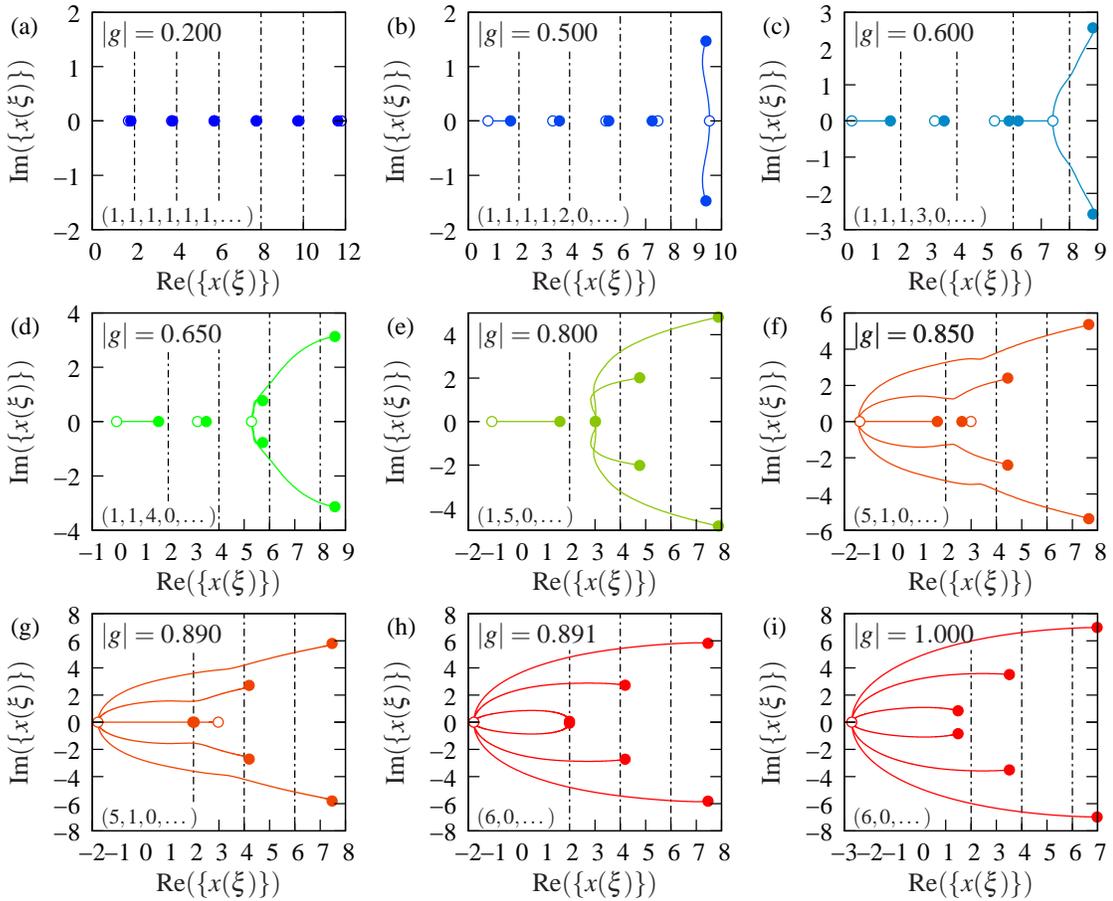}
\caption{The path of the deformed RG variables $\{x(\xi)\}$ in the complex plain for the  picket-fence model of Sec.\ \ref{subsection:tda} with $|g|=0.200$ (a), $|g|=0.500$ (b), $|g|=0.600$ (c), $|g|=0.650$ (d), $|g|=0.800$ (e), $|g|=0.850$ (f), $|g|=0.890$ (g), $|g|=0.891$ (h), and $|g|=1.000$ (i).  The path is given starting from the exact RG variables $\{x(1)\}$ (filled circles) and ending at the bosonic eigenmodes $\{x(0)\}=\{\hbar\omega\}$ (open circles).  The connecting multiphonon distribution $(\nu_1,\nu_2,\dots,\nu_m)$ is included in each figure, and the single-particle levels at $2\varepsilon_i$ ($i=1\dots m$) are also given in dashed-dotted lines as reference points for the multiphonon states.}\label{figure:tda2rg}
\end{figure*}

The previous analysis demonstrates the behavior of the RG variables as they move away from the bosonic contraction limit.  However, the description (\ref{bosonic:deformedrg:rgvariablesnearcontractionlimit}) is only valid in the vicinity of the contraction limit, and it is impossible in general to extend this to the fully restored quasispin limit $\xi=1$.  Nonetheless, it is known that the RG variables can be related to the roots of the associated Laguerre polynomials in both the weak- and strong-coupling limit \cite{ortiz:2005,yuzbashyan:2007}.  It will be demonstrated in Appendix \ref{appendix:limitsdeformedrg:strongcoupling} that this result is extendible along the full path $\xi\in[0,1]$, allowing for a 1-to-1 adiabatic connection between the exact ground state and its bosonic counterpart in the contraction limit.  The RG variables are given by
\begin{equation}\label{bosonic:deformedrg:rgvariablesstrongcoupling}
x_\alpha(\xi)=-g\xi y_{l(\alpha)} + \tfrac{\sum_{i=1}^m\xi d_i(\xi)2\varepsilon_i}{\xi d(\xi)}+\mathcal{O}(\tfrac{1}{g}),
\end{equation}
with $y_{l}$ shorthand for the $l$th root of the associated Laguerre polynomial $L^{-2d(\xi)-1}_N(y)$, and $d(\xi)=\sum_{i=1}^m d_i(\xi)$.  The value $-2d(\xi)-1$ is negative, so all roots are non-real, except for at most one real root when $N$ is odd \cite{szego:1975,shastry:2001}.   The description for the RG variables (\ref{bosonic:deformedrg:rgvariablesstrongcoupling}) is in principle valid for any finite value of $\xi$.  Caution is required for the contraction limit $\xi\rightarrow0$, as the parameter $-2d(\xi)-1$ of the associated Laguerre polynomial becomes singular.  It can be shown that the roots of the associated Laguerre polynomial $L^k_n(y)$ with $|k|\rightarrow\infty$ behave approximately like \cite{calogero:1978,temme:1990}
\begin{equation}
y_l = k+\sqrt{2k} z_{n,l},
\end{equation}
with $z_{n,l}$ the $l$th root of the Hermite polynomial $H_n(z)$.  Using this relation in (\ref{bosonic:deformedrg:rgvariablesstrongcoupling}), combined with the fact that $\xi d_i(\xi)=\frac{1}{4}\Omega_i-\xi\frac{1}{2}v_i$ is well-defined and finite, we obtain the near-contraction limit for the ground state of the strong-coupling regime
\begin{equation}
x_\alpha(\xi)=g\left[\tfrac{1}{2}\Omega-i\sqrt{\xi\Omega}z_{N,l(\alpha)}\right]+\tfrac{\sum_{i=1}^m\Omega_i2\varepsilon_i}{\Omega}+\mathcal{O}(\xi,\tfrac{1}{g}),
\end{equation}
with $\Omega=\sum_{i=1}^m\Omega_i$.  This result is consistent with (\ref{bosonic:deformedrg:rgvariablesnearcontractionlimit}), as the connecting multiphonon distribution of the ground state is given by $(N,0,\dots,0)$, with lowest (collective) eigenmode 
\begin{equation}
\hbar\omega_1=g\tfrac{1}{2}\Omega+\tfrac{\sum_{i=1}^m\Omega_i2\varepsilon_i}{\Omega}+\mathcal{O}(\tfrac{1}{g}).
\end{equation}
The analysis of the strong-coupling regime within the HS formalism shows that the exact ground state of the reduced BCS Hamiltonian can be unambiguously related to the $(N,0,\dots,0)$ collective multiphonon state by adiabatically switching off the Pauli exclusion principle via a pseudo deformation of the quasi spin.  It should be mentioned that the HS correspondence allows for a qualitative description, which is only correct within the validity domain of the approximation.  For an exact assessment of how the exact exact states are related to the bosonic counterpart, it is necessary to numerically solve the deformed RG equations (\ref{bosonic:deformedrg:rgequations}).  This was done for the picket-fence model \cite{sambataro:2007}, described in the previous section, for a selected set of coupling constants $g$.  The results are presented in Fig.\ \ref{figure:laguerrehermite} and Fig.\ \ref{figure:tda2rg}.  Fig.\ \ref{figure:laguerrehermite} shows the path of the RG variables in the complex plane as the pseudo deformation $\xi$ is tuned down from $1$ to $0$.  
This is done for a system in the intermediate- ($|g|=0.650$ in Fig.\ \ref{figure:laguerrehermite}a) and strong-coupling ($|g|=5.000$ in Fig.\ \ref{figure:laguerrehermite}b) regime.  It is readily observed that the RG variables evolve towards real and degenerate points in the complex plane, associated with the multiphonon states.  For the intermediate case, the associated multiphonon state is $(1,1,4,0,\dots)$, whereas the strong-coupling case connects with the collective $(6,0,0,\dots)$ multiphonon state.  In both cases, the exact path has been confronted with the predictions of the near-contraction limit, up to $\mathcal{O(\xi)}$ (see Appendix \ref{appendix:limitsdeformedrg:nearcontraction} on how the approximate values of Eq.\ (\ref{bosonic:deformedrg:rgvariablesnearcontractionlimit}) can be improved up to $\mathcal{O(\xi)}$).  It can be concluded from the insets in Fig.\ \ref{figure:laguerrehermite} that the near-contraction limit of the deformed RG equations is  well understood using the HS correspondence.  In addition, the results for the strong-coupling case in Fig.\ \ref{figure:laguerrehermite}b are compared with the approximate expression (\ref{bosonic:deformedrg:rgvariablesstrongcoupling}) in the strong-coupling limit.  The general trend of the RG equations in the complex plane is rather well reproduced, taking into account that the errors are expected to be of order $\mathcal{O}(g^{-2})$ with respect to the order $\mathcal{O}(g)$ of the RG variables.  So, it is confirmed that the RG variables in the strong-coupling regime indeed evolve towards a fully collective multiphonon state, as anticipated.  

It can be inferred from Fig.\ \ref{figure:laguerrehermite} that the exact ground state of the reduced BCS Hamiltonian can connect to different multiphonon states for different coupling constants $g$.  This has been investigated in detail for the ground state of the picket fence model of section \ref{subsection:tda} where the interaction strength has been varied from $g=-2.000$ to $g=0.000$.  The results of this study can be found in Figures \ref{figure:tda2rg} \& \ref{figure:sambataro:rgvariables}.
It was observed that the exact ground state in the strong-coupling regime is always connected with the $(6,0,\dots,0)$ multiphonon state (see Fig.\ \ref{figure:laguerrehermite} and Fig.\ \ref{figure:tda2rg}i for respectively $|g|=5.000$ and $|g|=1.000$), until the critical interaction strength $|g_c|=0.890$ was reached as a lower limit.  Crossing this critical point gave rise to a change in multiphonon character in the contraction limit.  Instead of the fully collective state, the exact state chose to connect with the $(5,1,0,\dots,0)$ multiphonon state (see Fig.\ \ref{figure:tda2rg}f) for $|g|=0.850$).  This can be understood by realizing that one of the complex conjugate pair of RG variables changes into two real-valued variables at the critical interaction strength.  In order to show the subtleties involved with the change of the multiphonon state around the critical interaction strength $g$, the path towards the contraction limit was plotted in Fig.\ \ref{figure:tda2rg}g and Fig.\ \ref{figure:tda2rg}h for $|g|=0.890$ and $|g|=0.891$ respectively.  The resolution of Fig.\ \ref{figure:tda2rg} is hardly sufficient to resolve the two critical RG variables in both the figures, however the path to the multiphonon states is clearly different.  Remarkably, the next $g$ where the multiphonon state changes character is not at the next critical interaction strength $|g_c|=0.603$.  It was observed that the next two instances occurred at $|g|=0.845$ and $|g|=0.719$ where the multiphone state changes to $(1,5,0,\dots,0)$ and $(1,1,4,0,\dots,0)$ respectively, which can be interpreted as a simple rearrangement of the two real-valued and four complex-valued RG variables into a less collective configuration.  The path towards the $(1,5,0,\dots,0)$ and $(1,1,4,0,\dots,0)$ multiphonon states are illustrated for $|g|=0.800$ and $|g|=0.650$ in Fig.\ \ref{figure:tda2rg}d and Fig.\ \ref{figure:tda2rg}e.  The near-contraction limit of the latter figure has already been discussed in detail in Fig.\ \ref{figure:laguerrehermite}a.  The next point where the multiphonon configuration alters is at the critical point $|g_c|=0.603$, where the $(1,1,4,0,\dots,0)$ changes into $(1,1,1,3,0,\dots,0)$, which can again be related to the breaking of a complex conjugate pair of RG variables into two real-valued RG variables.  The path to the $(1,1,1,3,0,\dots,0)$ multiphonon state is illustrated for $|g|=0.600$ in Fig.\ \ref{figure:tda2rg}c.  Next, there is again a change in multiphonon distribution related to a rearrangement of the real- and complex-valued RG variables at $|g|=0.580$ to a less collective $(1,1,1,1,2,0\dots,0)$ (see Fig.\ \ref{figure:tda2rg}b).  Finally, the last change occurs at the critical value $|g_c|=0.393$, below which each exact state corresponds to a $(1,1,1,1,1,1,0,\dots,0)$ multiphonon state (as illustrated for $|g|=0.200$ in Fig.\ \ref{figure:tda2rg}a).  The physical interpretation of this configuration is a simple filling of the single-particle levels with pairs until the Fermi level is filled.  The situation of the picket-fence model of subsection \ref{subsection:tda} is summarized in Fig.\ \ref{figure:sambataro:rgvariables} and Table \ref{table:multiphononchanges}.
\begin{table}[!htb]
\begin{tabular}{rl|cccccccc}
\hline
 $|g|$ & & $\nu_1$ & $\nu_2$ & $\nu_3$ &  $\nu_4$ & $\nu_5$ & $\nu_6$ & $\dots$ & $\nu_{12}$\\
\hline
\hline
  $0.000$ &     & 1 & 1 & 1 & 1 & 1 & 1 & $\dots$ & 0 \\
  $0.393$ & $c$ & 1 & 1 & 1 & 1 & 2 & 0 & $\dots$ & 0 \\
  $0.580$ & $r$ & 1 & 1 & 1 & 3 & 0 & 0 & $\dots$ & 0 \\
  $0.603$ & $c$ & 1 & 1 & 4 & 0 & 0 & 0 & $\dots$ & 0 \\
  $0.719$ & $r$ & 1 & 5 & 0 & 0 & 0 & 0 & $\dots$ & 0 \\
  $0.845$ & $r$ & 5 & 1 & 0 & 0 & 0 & 0 & $\dots$ & 0 \\
  $0.891$ & $c$ & 6 & 0 & 0 & 0 & 0 & 0 & $\dots$ & 0 \\
\hline
\end{tabular}
\caption{The coupling strengths $|g|$ where the connecting multiphonon character changes for the picket-fence model of Sec.\ \ref{subsection:tda}.  The given multiphonon distributions $(\nu_1,\nu_2,\dots,\nu_m)$ in the row of each $|g|$ is the distribution into which the change occurs for an increasing $|g|$.  The label $r$ and $c$ denote whether the coupling constant corresponds to respectively a \emph{r}earrangement or a \emph{c}ritical RG variable.  These interaction strengths are visualized in Fig.\ \ref{figure:sambataro:rgvariables}.}\label{table:multiphononchanges}
\end{table}
The coupling constants for which the connecting multiphonon distribution changes have been denoted by dotted and dashed lines in Fig.\ \ref{figure:sambataro:rgvariables}, with the distinction that the dashed lines denote the critical interaction strengths associated with a recombination of RG variables into complex conjugate pairs, whereas the dotted lines indicate the values of $g$ where the connecting multiphonon changes without such a recombination.  The exact values where the multiphonon distributions change are given in Table \ref{table:multiphononchanges} up to three decimal places.

Now, it would be interesting to put the present classification of the ground state into different multiphonon distributions against the overlaps of the exact ground state with the $pp$-TDA multiphonon states $\langle\psi\{x\}|\psi\{\hbar\omega\}\rangle$ of Sec.\ \ref{subsection:tda} (see Fig.\ \ref{figure:sambataro:overlaps}).  The present discussion answers the question of Sec.\ \ref{subsection:tda} whether it is possible to assign a change of multiphonon character to the critical point $|g_c|=0.603$.  However, caution is required identifying the contracted multiphonon states (\ref{bosonic:deformedrg:betheansatzstatecontractionlimit}) with the $pp$-TDA multiphonon states (\ref{bosonic:tda:multiphononstate}).  For the sake of comparison, the overlaps of the TDA phonon states with the exact ground state for the distributions given in Table \ref{table:multiphononchanges} have been plotted as well in Fig.\ \ref{figure:sambataro:overlaps}.  It can be seen that the identification (contraction-limit vs.\ $pp$-TDA multiphonon states) works rather well in the weak- and strong-coupling regime.  Indeed, for what the weak-coupling regime is concerned, the domains where the exact ground state connects with the multiphonon states $(1,1,1,1,1,1,0,\dots,0)$ and $(1,1,1,1,2,0,\dots,0)$ correspond largely with the domains where the $(1,1,1,1,1,1,0,\dots,0)$ and $(1,1,1,1,2,0,\dots,0)$ $pp$-TDA multiphonon states give the maximum overlap (compare Table \ref{table:multiphononchanges} with Fig.\ \ref{figure:sambataro:overlaps}).  Similarly, in the strong-coupling regime, there is a correspondence between the domains of the connecting multiphonon states $(5,1,0,\dots,0)$ and $(6,0,\dots,0)$ with respectively the domains where the $pp$-TDA states with eigenmode distribution $(5,1,0,\dots,0)$ and $(6,0,\dots,0)$ give the maximum overlap.  However, the connection is not unambiguous in the intermediate regime.  Closer inspection of Fig.\ \ref{figure:sambataro:overlaps} reveals that the overlaps of the $pp$-TDA states with eigenmode distribution $(1,1,1,3,0,\dots,0)$, $(1,1,4,0,\dots,0)$ and $(1,5,0,\dots,0)$ (\emph{cfr}.\ Table \ref{table:multiphononchanges}) may grow relatively large in the intermediate regime, but they never result in the maximum overlap for a given interaction strength.  This points again towards the intricate role the Pauli principle plays in the intermediate regime.  The exclusion principle is fully active in the calculation of the overlaps of the exact with the $pp$-TDA multiphonon states, because both states are living in the same Hilbert space.  So, the TDA state with maximum overlap will be the result of a subtle trade off of constructing a highly collective multiphonon state (\emph{i.e.}\ more phonons with eigenmode $\hbar\omega_1$) against the action of the Pauli principle,  washing away part of the collectivity.  Within the framework of the pseudo-deformed quasispin algebra, this trade-off is irrelevant as the Pauli principle is totally suppressed in the contraction limit.

\begin{figure}[!htb]
 \includegraphics{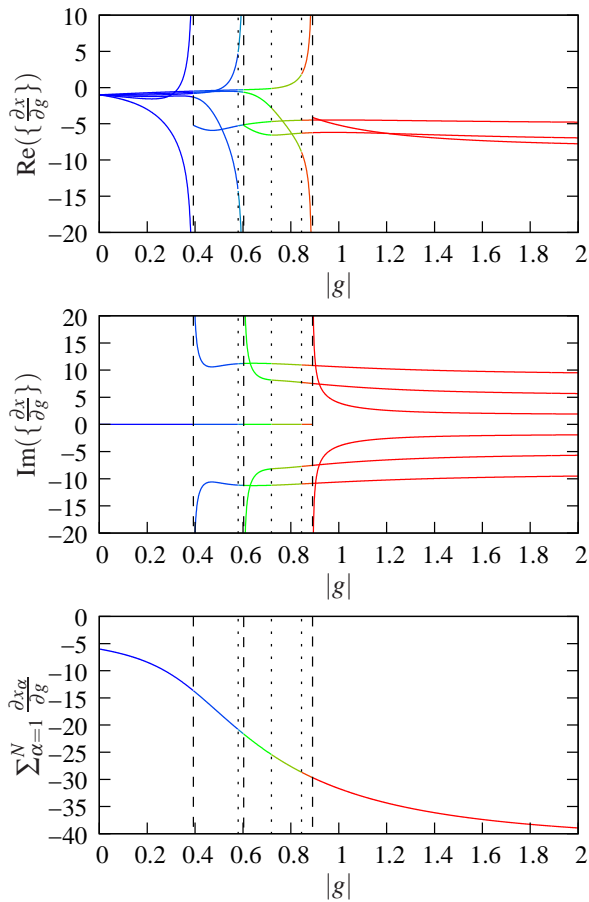}
 \caption{The real part (a), imaginary part (b), and the total sum (c) of the derivatives of the exact RG variables with respect to the interaction strength $g$.  The singular interaction strengths are highlighted by dotted and dashed lines, where dashed lines denote critical values $|g_c|$ and dotted lines highlight the values of $|g|$ where a multiphonon rearrangement occurs (see Fig.\ \ref{figure:sambataro:rgvariables} and Table \ref{table:multiphononchanges}).  }\label{figure:derivatives}
\end{figure}

The critical values $g_c$ are clearly related to a change of multiphonon character in the contraction limit.  However, as is clear from the previous discussion, in addition to these critical values, a few other values of $g$ have been found where the connecting multiphonon state changes character.  It would be interesting to investigate a possible relation between these changes in multiphonon character and higher-order singular points in the RG variables.  For this purpose, the derivatives of the RG variables $\{\frac{\partial x}{\partial g}\}$ have been plotted in Fig.\ \ref{figure:derivatives}, with the anticipation that some additional structure could be found at these particular values of $g$.
Unfortunately, as can be seen from the figure, no hints of a possible higher-order singular point could be found at the mentioned values of $g$.  Therefore, the nature of these additional points remains largely unclear.  It is worth mentioning that none of these points were found in previous analysis' within the pseudo-deformed quasispin formalism \cite{debaerdemacker:2011,debaerdemacker:2012}, including a study of the numerical test-case model of Rombouts \emph{et.\ al.} \cite{rombouts:2004}.

The discussion has focused thusfar on connecting the exact RG states (\ref{rg:betheansatzstate}) with the corresponding contracted multiphonon state (\ref{bosonic:deformedrg:betheansatzstatecontractionlimit}).  A valuable question would be how the connection works in the opposite direction, \emph{i.e.}\ given a multiphonon state (\ref{bosonic:deformedrg:betheansatzstatecontractionlimit}), is there a \emph{physical} state (\ref{rg:betheansatzstate}), corresponding to an exact eigenstate of the reduced BCS Hamiltonian (\ref{rg:hamiltonian})?  The Hilbert space $\mathcal{H}_{\xi=0}$ spanned by all possible multiphonon states is much larger than the Hilbert space $\mathcal{H}_{\xi=1}$ spanned by all possible hard-core bosonic states.  For instance, the dimensions of a system of $N$ pairs living in  $m$ twofold degenerate ($\Omega_i=2$) single-particle levels are given by
\begin{equation}
\dim{\mathcal{H}}_{\xi=0}=\left(\begin{array}{c}m+N-1\\N\end{array}\right),\quad\dim{\mathcal{H}} _{\xi=1}=\left(\begin{array}{c}m\\N\end{array}\right).
\end{equation}
Therefore, we may expect that there are multiphonon states for which there is no corresponding exact state.  Indeed, as an example, adiabatically turning on the Pauli principle starting from the collective $(6,0,\dots,0)$ multiphonon state did not result into an exact eigenstate at $\xi=1$ for $|g|<0.891$.  The solution broke down at a given $\xi<1$ along the path.  This is somewhat unfortunate, because the quasi-deformed algebra approach could have the potential to act as a convenient RG solver, complementary to other approaches \cite{rombouts:2004,dominguez:2006,dussel:2007,faribault:2011,marquette:2012}.  Multiphonon states are considerably more tractable from a numerical point of view than the RG equations, and the Pauli principle can be reintroduced in a very controlled fashion.  The fact that the Hilbert space of the genuine bosons is larger than the Hilbert space of the fermionic pairs, would imply that many multiphonon states may be tried before the corresponding exact state is found. Nevertheless, as has been shown in the present manuscript, as well as in previous studies \cite{debaerdemacker:2011,debaerdemacker:2012}, physical insight and intuition can prove very helpful in finding the multiphonon states corresponding to the exact ground state and excited states \cite{sambataro:2007,debaerdemacker:2011}.  As a matter of fact, the exact ground states in the present work have been obtained using this idea of reconstruction the exact ground state from the connection multiphonon state.
\section{Conclusions}\label{section:conclusions}
In conclusion, an adiabatic connection has been made between the hard-core bosonic eigenstates of the reduced BCS Hamiltonian and the multiphonon states characteristic to bosonic approximations, using a pseudo deformation of the $su(2)$ quasispin algebra.  The pseudo deformation is not a genuine deformation because it can be transformed into an effective quasi-spin algebra with rescaled irrep labels.  Only in the contraction limit of the deformation parameter transforms the deformed $su(2)$ into a Heisenberg-Weyl algebra $hw(1)$, the fundamental spectrum generating algebra of bosonic excitations.   It is shown that the reduced BCS Hamiltonian remains Richardson-Gaudin integrable along the deformation path of the pseudo-deformed quasispin algebra with a corresponding set of deformed Richardson-Gaudin equations.  This set of deformed Richardson-Gaudin equations become equivalent to the seniority-free secular equation of the $pp$-Tamm-Dancoff Approximation.  The near-contraction and strong-coupling limit of the deformed Richardson-Gaudin equations have been investigated using a Heine-Stieltjes correspondence\, and an example using a so-called picket-fence model is discussed in detail.  The present approach allowed for an interpretation of the critical interaction strengths $g_c$ as those points where the connecting multiphonon states change collective nature.  Also, the potential of the adiabatic connection as a Richardson-Gaudin solver has been briefly discussed, complementary to recent developments in the numerics of the Richardson-Gaudin equations.  Finally, it would be interesting to investigate how this pseudo-deformation scheme is positioned within the huge realm of bosonic mappings \cite{klein:1991} and bosonic approximations to pairing dynamics of fermionic systems \cite{iachello:1987}.
\begin{acknowledgments}
The author is an ''FWO Vlaanderen'' post-doctoral fellow and acknowledges two FWO travel grants for a ''long stay abroad'' at the University of Toronto and the University of Notre Dame.  It is my pleasure to acknowledge Piet Van Isacker (GANIL) and Stefan Rombouts for triggering my interest into integrable systems, and for the many discussion that followed since then.  My gratitude also goes to the mathematical physics group of David J.\ Rowe at the University of Toronto for the many insightful discussions during the initial stages of this work.  Dimitri Van Neck (Ghent University), Paul Ayers and Paul Johnson (McMaster University) are acknowledged for taking this work already one step ahead of me.  Finally, I would like to thank Kris Heyde (Ghent University), John Wood (Georgia Institute of Technology) and Veerle Hellemans (Universit\'e Libre de Bruxelles) for continued interest in this work and support.
\end{acknowledgments}
\appendix

\section{Deriving the RG equations}\label{appendix:derivationrg}
It is straightforward to derive the RG equations using a commutator scheme \cite{rowe:2010,ortiz:2005}.  In this scheme, the Hamiltonian (\ref{rg:hamiltonian}) (or Eq.\ (\ref{bosonic:deformedrg:hamiltonian}) in the deformed case) is commuted through the product state (\ref{rg:betheansatzstate}) (or Eq.\ (\ref{bosonic:deformedrg:betheansatzstate})) until it acts on the pair vacuum.  The RG equations are obtained by requiring the non-diagonal by-products to vanish.   It is convenient to introduce the following notation.  Let $\hat{O}_i$ be a level-dependent operator with the Roman index $i=1\dots m$ denoting the level.  The associated Gaudin operator is given by
\begin{equation}
\hat{O}_\alpha=\sum_{i=1}^m\frac{\hat{O}_i}{2\varepsilon_i-x_\alpha}.
\end{equation}
with the Greek label $\alpha$ shorthand for the RG variable $x_\alpha$.  Further, it is also convenient for notational purposes to introduce the operator $\hat{O}=\sum_{i=1}^m\hat{O}_i$.  Using this notation, the Schr\"odinger equation becomes
\begin{equation}
\hat{H}(\xi)\prod_{\alpha=1}^N\hat{S}^\dag_\alpha(\xi)|\theta\rangle=E(\xi)\prod_{\alpha=1}^N\hat{S}^\dag_\alpha(\xi)|\theta\rangle.
\end{equation}
The general deformed case (as defined in Sec.\ \ref{subsection:quasispin}) will be used in the present Appendix, as the regular RG formalism can be obtained from the $\xi=1$ limit. By commuting the Hamiltonian through the product state, we derive the following expression
\begin{align}\label{derivationrg:commutatorscheme}
\hat{H}(\xi)&\prod_{\alpha=1}^N\hat{S}^\dag_\alpha(\xi)|\theta\rangle=\prod_{\alpha=1}^N\hat{S}^\dag_\alpha(\xi)\hat{H}(\xi)|\theta\rangle\\
&+\sum_{\beta=1}^N\prod_{\alpha\neq\beta}^N \hat{S}^\dag_\alpha(\xi)[\hat{H}(\xi),\hat{S}^\dag_\beta(\xi)]|\theta\rangle\notag\\
&+\sum_{\beta=1}^N\sum_{\gamma=\beta+1}^N\prod_{\alpha\neq\beta,\gamma}^N \hat{S}^\dag_\alpha(\xi)[[\hat{H}(\xi),\hat{S}^\dag_\beta(\xi)],\hat{S}^\dag_\gamma(\xi)]|\theta\rangle\notag,
\end{align}
where the property that the double commutator with the Hamiltonian (\ref{bosonic:deformedrg:hamiltonian})  commutes with all creation operators has been used in the last line.  The single and double commutators can now be calculated explicitly using the deformed algebra (\ref{bosonic:quasispin:deformedalgebra}).  
\begin{align}
&[\hat{H}(\xi),\hat{S}^\dag_\beta(\xi)]=x_\beta\hat{S}^\dag_\beta(\xi)\\
&\qquad+\hat{S}^\dag(\xi)\left[1-g\left(2\xi\hat{S}^0_\beta(\xi)+(\xi-1)\tfrac{1}{2}\Omega_\beta\right)\right]\notag\\
&[[\hat{H}(\xi),\hat{S}^\dag_\beta(\xi)],\hat{S}^\dag_\gamma(\xi)]=-2g\xi\hat{S}^\dag(\xi)\frac{\hat{S}^\dag_\beta(\xi)-\hat{S}^\dag_\gamma(\xi)}{x_\beta-x_\gamma}
\end{align}
The action of these commutators on the pair vacuum state simplifies Eq.\ (\ref{derivationrg:commutatorscheme}) to
\begin{align}
\hat{H}(\xi)&\prod_{\alpha=1}^N\hat{S}^\dag_\alpha(\xi)|\theta\rangle=\left[\sum_{\beta=1}^Nx_\beta+\sum_{i=1}^m\varepsilon_iv_i\right]\prod_{\alpha=1}^N\hat{S}^\dag_\alpha(\xi)|\theta\rangle\notag\\
&+\sum_{\beta=1}^N\left[1+2g\xi d_\beta(\xi)-2g\xi\sum_{\gamma\neq\beta}^N\frac{1}{x_\gamma-x_\beta}\right]\notag\\
&\qquad\times\prod_{\alpha\neq\beta}^N \hat{S}^\dag_\alpha(\xi)\hat{S}^\dag(\xi)|\theta\rangle,
\end{align}
taking into account that $2\xi d_\alpha+(\xi-1)\frac{1}{2}\Omega_\alpha=2\xi d_\alpha(\xi)$ (see Eq.\ (\ref{bosonic:quasispin:deformedirreplabel})).  The crucial observation is now that the product state $\prod_{\alpha=1}^N\hat{S}^\dag_\alpha|\theta\rangle$ will only be an eigenstate iff the non-diagonal terms vanish identically.  This is the case if 
\begin{equation}
1+2g\xi d_\beta(\xi)-2g\xi\sum_{\gamma\neq\beta}^N\frac{1}{x_\gamma-x_\beta}=0,\quad\forall\beta=1\dots N,
\end{equation}
which constitutes identically the set of deformed RG equations (\ref{bosonic:deformedrg:rgequations}).  It is worth pointing out that the present general procedure logically applies to the regular RG formalism for the BCS Hamiltonian ($\xi=1$) as to the bosonic contraction limit ($\xi=0$).  Moreover, it has been used \cite{tsyplyatyev:2010} to derive the Bethe Ansatz of the Dicke model, describing a mixture of hard-core bosonic and genuinely bosonic pairs.

\section{Limits of the deformed RG equations}\label{appendix:limitsdeformedrg}
In some limiting cases, the solutions of the RG equations can be related to the roots of special functions, using the Heine-Stieltjes correspondence \cite{stieltjes:1885,szego:1975}.  This will be discussed in more detail in the present Appendix for the near-contraction limit and the strong-coupling regime over the whole regime of the deformation parameter.

\subsection{The near-contraction limit}\label{appendix:limitsdeformedrg:nearcontraction}
Assume we have a multiphonon state (\ref{bosonic:deformedrg:betheansatzstatecontractionlimit}) in the contraction limit, with a given eigenmode distribution $(\nu_1,\nu_2,\dots,\nu_m)$.  The RG variables in this limit are given by $x_\alpha(0)=\hbar\omega_{k(\alpha)}$, with $\hbar\omega_k$ the solutions of the secular equation (\ref{bosonic:deformedrg:rgequationscontractionlimit}) and $k(\alpha)$ an integer-valued function picking the $k$th eigenmode for the $\alpha$th pair.  Making use of (\ref{bosonic:quasispin:deformedirreplabel}), the deformed RG equations (\ref{bosonic:deformedrg:rgequations}) can be written explicitly as
\begin{equation}\label{limitsdeformedrg:rgequations}
1-g\sum_{i=1}^m\frac{\xi v_i-\frac{1}{2}\Omega_i}{2\varepsilon_i-x_\alpha}-2g\xi\sum_{\beta\neq\alpha}^N\frac{1}{x_\beta-x_{\alpha}}=0.
\end{equation}
Assume that the RG variables in the near-contraction limit are given by
\begin{equation}
x_\alpha(\xi)=\hbar\omega_{k(\alpha)}+\sqrt{\xi}x^{(1)}_\alpha+\mathcal{O}(\xi).
\end{equation}
We need to make a distinction in the summation over $\beta$ in equation (\ref{limitsdeformedrg:rgequations}) between the pairs $x_\beta(\xi)$ with the same eigenmode as $x_\alpha(\xi)$ ($k(\beta)=k(\alpha)$) and the pairs $x_{\beta^\prime}(\xi)$ with a different eigenmode ($k(\beta^\prime)\neq k(\alpha)$)
\begin{align}
 1-&g\sum_{i=1}^m\frac{\xi v_i-\tfrac{1}{2}\Omega_i}{2\varepsilon_i-\hbar\omega_{k(\alpha)}-\sqrt{\xi}x^{(1)}_\alpha}\\
&-2g\xi\sum_{\substack{\beta^\prime\neq\alpha\\k(\beta^\prime)\neq k(\alpha)}}\frac{1}{\hbar\omega_{k(\beta^\prime)}-\hbar\omega_{k(\alpha)}+\sqrt{\xi}(x^{(1)}_{\beta^\prime}-x^{(1)}_\alpha)}\notag\\
&-2g\sqrt{\xi}\sum_{\substack{\beta\neq\alpha\\k(\beta)=k(\alpha)}}\frac{1}{x^{(1)}_\beta-x^{(1)}_\alpha}=0,\qquad(\forall \alpha=1..N)\notag.
\end{align}
Now, $\xi$ can be chosen small enough for a meaningful series expansion up to first order in $\sqrt{\xi}$.
\begin{align}
 1+&\frac{g}{2}\sum_{i=1}^m\frac{\Omega_i}{2\varepsilon_i-\hbar\omega_{k(\alpha)}}\\
&+g\sqrt{\xi}\left[\frac{1}{2}\sum_{i=1}^m\frac{\Omega_i}{(2\varepsilon_i-\hbar\omega_{k(\alpha)})^2}x^{(1)}_\alpha-\sum_{\beta\neq\alpha}\frac{2}{x^{(1)}_\beta-x^{(1)}_\alpha}\right]\notag\\
&+\mathcal{O}(\xi)\approx0,\qquad(\forall \alpha=1\dots N),\notag
\end{align}
with the summation over $\beta$ limited to those pairs for which $k(\beta)=k(\alpha)$.  The key observation is that the terms in $\mathcal{O}(1)$ vanish by definition because $\hbar\omega_k$ has been chosen as such that it resolves the bosonic secular equation (\ref{bosonic:deformedrg:rgequationscontractionlimit}).  As a result, we obtain the Stieltjes equations \cite{shastry:2001,stieltjes:1914}
\begin{equation}\label{limitsdeformedrg:stieltjes}
a_{k(\alpha)}x^{(1)}_\alpha=\sum_{\beta\neq\alpha}\frac{2}{x^{(1)}_\beta-x^{(1)}_\alpha},\quad\forall\alpha=1\dots N
\end{equation}
with $a_l=\frac{1}{2}\sum_{i=1}^m\frac{\Omega_i}{(2\varepsilon_i-\hbar\omega_{l})^2}$.  Because the summation $\beta$ runs only over the RG variables with $k(\beta)=k(\alpha)$, the full $N$-pair problem breaks down into separate blocks of RG variables with equal $x_{\alpha}=\hbar\omega_{k(\alpha)}$, for which the Stieltjes equations can be solved independently from each other.  This can be done using the Heine-Stieltjes connection \cite{shastry:2001,stieltjes:1914}, of which the main ideas are summarized below.  Assume we want to solve the Stieltjes equations for $\nu$ variables $x^{(1)}_\alpha$ ($\alpha=1\dots\nu$).  Let's introduce the Heine-Stieltjes polynomial,
\begin{equation}
 P(x)=\prod_{\alpha=1}^\nu(x-x^{(1)}_\alpha),
\end{equation}
defined as the polynomial of degree $\nu$ with the solution of (\ref{limitsdeformedrg:stieltjes}) as the roots.  It is straightforward to derive that
\begin{equation}
\frac{P^{\prime\prime}(x^{(1)}_\alpha)}{P^{\prime}(x^{(1)}_\alpha)}=2\sum_{\beta\neq\alpha}\frac{1}{x^{(1)}_\alpha-x^{(1)}_\beta},\quad(\forall\alpha=1\dots\nu)
\end{equation}
As a result, the Stieltjes equations (\ref{limitsdeformedrg:stieltjes}) can be rewritten as
\begin{equation}\label{limitsdeformedrg:stieltjesrewritten}
a_kx^{(1)}_\alpha P^\prime(x^{(1)}_\alpha)+P^{\prime\prime}(x^{(1)}_\alpha)=0, \quad(\forall\alpha=1\dots\nu)
\end{equation}
Now, we can define the polynomial
\begin{equation}
Q(x)=a_kxP^\prime(x)+P^{\prime\prime}(x),
\end{equation}
which has the same degree $\nu$ as $P(x)$, and the same set of roots (see equation (\ref{limitsdeformedrg:stieltjesrewritten})), so $Q(x)$ must be a multiple of $P(x)$.  It is straightforward to see from the definition of $Q(x)$ that the multiplication factor is $a_k\nu$.  Therefore, we obtain the equation
\begin{equation}
a_kP^\prime(x)+P^{\prime\prime}(x)=a_k\nu P(x).
\end{equation}
A change of variable $z=-i\sqrt{a_k/2}x$ transforms this equation into the Hermite differential equation \cite{arfken:2001}
\begin{equation}
H^{\prime\prime}(z)-2zH^{\prime}(z)+2\nu H(z)=0. 
\end{equation}
In conclusion, the solution of the Stieltjes equations (\ref{limitsdeformedrg:stieltjes}) can be related to the roots $\{z_{\nu}\}$ of the Hermite polynomials $H_\nu(z)$.  Consequently, the RG variables in the near-contraction limit are given by
\begin{equation}\label{limitsdeformedrg:stieltjes:rgvariable}
x_\alpha(\xi)=\hbar\omega_{k(\alpha)}+i\sqrt{\tfrac{2\xi}{a_{k(\alpha)}}}z_{\nu_{k(\alpha)},l(\alpha)}+\mathcal{O}(\xi),
\end{equation}
with $z_{\nu,l}$ the $l$th root of the Hermite polynomial $H_\nu(z)$.

It is understood from equation (\ref{limitsdeformedrg:stieltjes:rgvariable}) that the first-order correction in $\sqrt{\xi}$ is purely imaginary.   Therefore, if a qualitative description of the real part of the RG variables near the contraction limit beyond $\mathcal{O}(1)$ is desired, it is necessary to include higher-order terms in the approximate solution of the deformed RG equations (\ref{limitsdeformedrg:rgequations})
\begin{equation}
x_\alpha(\xi)=\hbar\omega_{k(\alpha)}+\sqrt{\xi}x^{(1)}_\alpha+\xi x^{(2)}_\alpha+ \mathcal{O}(\xi^{\frac{3}{2}}).
\end{equation}
In addition to the Stieltjes equations at order $\mathcal{O}(\sqrt{\xi})$, we obtain a linear set of equations for the correction variables $\{x^{(2)}\}$ at order $\mathcal{O}(\xi)$
\begin{align}
&a_{k(\alpha)}x^{(2)}_\alpha+\sum_{\substack{\beta\neq\alpha\\k(\beta)\neq k(\alpha)}} \frac{2(x^{(2)}_\beta-x^{(2)}_\alpha)}{(x^{(1)}_\beta-x^{(1)}_\alpha)^2}\notag\\
&\quad=-b_{k(\alpha)}(x^{(1)})^2+c_{k(\alpha)}+B(k(\alpha)),
\end{align}
with $a_l$ defined previously, $b_l=\frac{1}{2}\sum_{i=1}^m\frac{\Omega_i}{(2\varepsilon_i-\hbar\omega_l)^3}$, and $c_l=\sum_{i=1}^m\frac{v_i}{2\varepsilon_i-\hbar\omega_l}$.   $B(k(\alpha))=\sum_{\beta^\prime\neq\alpha}\frac{1}{\hbar\omega_{k(\beta^\prime)}-\hbar\omega_{k(\alpha)}}$, with the summation $\beta^\prime$ restricted to the pairs for which $k(\beta^\prime)\neq k(\alpha)$, is a functional of the integer-valued function $k(\alpha)$.
It can be verified, taking the Stieltjes relation (\ref{limitsdeformedrg:stieltjes}) for the variables $\{x^{(1)}\}$ into account, that the solution to this set is given by the simple relation
\begin{align}
x^{(2)}_\alpha&=-\frac{b_{k(\alpha)}}{3a_{k(\alpha)}}(x^{(1)}_\alpha)^2+\frac{2b_{k(\alpha)}}{3a_{k(\alpha)}^2}(\nu_{k(\alpha)}-1)\notag \\
&\quad+\frac{c_{k(\alpha)}}{a_{k(\alpha)}}+\frac{B(k(\alpha))}{a_{k(\alpha)}}.
\end{align}
In conclusion, the $\mathcal{O}(\xi)$ corrections are purely real, and are directly related to the $\mathcal{O}(\sqrt{\xi})$ corrections.  These corrections have been taken into account for the comparison with exact results in Fig.\ \ref{figure:laguerrehermite}.

\subsection{The strong-coupling limit}\label{appendix:limitsdeformedrg:strongcoupling}
A similar analysis using the HS correspondence can be performed for the strong-coupling regime \cite{ortiz:2005,yuzbashyan:2007}.  The strong-coupling case of the present subsection is different with respect to the near-contraction case of the previous subsection as the present analysis is valid for any value of the pseudo deformation parameter $\xi$, whereas the previous analysis was only legitimate near the contraction limit.  It is instructive to start with the 1-pair problem, with the deformed RG equation (\ref{bosonic:deformedrg:rgequations}) given by
\begin{equation}
1-2g\sum_{i=1}^m\frac{\xi d_i(\xi)}{2\varepsilon_i-x_\alpha(\xi)}=0.
\end{equation}
The ground-state solution of this equation in the $|g|\rightarrow\infty$ limit is
\begin{equation}
x_\alpha(\xi)=2g\xi d(\xi) + \tfrac{\sum_{i=1}^m\xi d_i(\xi)2\varepsilon_i}{\xi d(\xi)} + \mathcal{O}(\tfrac{1}{g}), 
\end{equation}
with $d(\xi)=\sum_{i=1}^m d_i(\xi)$.  This result suggests that the ground state of the $N$-pair problem will be described by RG variables of the following form (the $\xi$ dependency will be omitted for notational simplicity)
\begin{equation}
x_\alpha=g x^{(0)}_\alpha + x^{(1)}_{\alpha} + \mathcal{O}(\tfrac{1}{g}),\quad\forall\alpha=1\dots N.
\end{equation}
Inserting this in the deformed RG equations (\ref{bosonic:deformedrg:rgequations}), we obtain
\begin{align}
 &1+2g\sum_{i=1}^m\frac{\xi d_i(\xi)}{2\varepsilon_i-gx^{(0)}_\alpha-x^{(1)}_\alpha}\\
&\quad-2g\xi\sum_{\beta=1,\neq\alpha}^N\frac{1}{g(x^{(0)}_\beta-x^{(0)}_\alpha)+x^{(1)}_\beta-x^{(1)}_\alpha}=0.
\end{align}
A Taylor expansion up to next-to-leading order simplifies this equation to
\begin{align}\label{limitsdeformedrg:rgequationstrongcouplingtaylor}
 &\left[1-2\frac{\xi d(\xi)}{x^{(0)}_\alpha}-2\xi\sum_{\beta=1,\neq\alpha}^N\frac{1}{x^{(0)}_\beta-x^{(0)}_\alpha}\right]\\
&+\frac{1}{g}\left[-\frac{\sum_{i=1}^m\xi d_i(\xi)2\varepsilon_i}{(x^{(0)}_\alpha)^2}+\frac{\xi d(\xi)x^{(1)}_\alpha}{(x^{(0)}_\alpha)^2}\right.\notag\\
&\qquad +\left.\xi\sum_{\beta=1,\neq\alpha}^N\frac{x^{(1)}_\beta-x^{(1)}_\alpha}{(x^{(0)}_\beta-x^{(0)}_\alpha)^2}\right]+\mathcal{O}(\tfrac{1}{g^2})=0,\forall\alpha.\notag
\end{align}
We require each order to vanish.  First, it can be verified that 
\begin{equation}
x_\alpha^{(1)}=\frac{\sum_{i=1}^m\xi d_i(\xi)2\varepsilon_i}{\xi d(\xi)},\quad\forall\alpha=1\dots N
\end{equation}
is a solution for the $\mathcal{O}(\frac{1}{g})$ term in equation (\ref{limitsdeformedrg:rgequationstrongcouplingtaylor}), independent from the solution $\{x^{(0)}\}$ of the $\mathcal{O}(1)$ term.  Second, the solution $\{x^{(0)}\}$ of the $\mathcal{O}(1)$ term can be found using the HS correspondence, similar to the near-contraction limit (Appendix \ref{appendix:limitsdeformedrg:nearcontraction}).  Again, by introducing the HS polynomial with the values $\{x^{(0)}\}$ as roots
\begin{equation}
P(x)=\prod_{\alpha=1}^N(x-x^{(0)}_\alpha), 
\end{equation}
it is possible to rewrite the equation resolving the $\mathcal{O}(1)$ term as
\begin{equation}
x^{(0)}_\alpha P^\prime(x^{(0)}_\alpha)-2\xi d(\xi)P^\prime(x^{(0)}_\alpha)+\xi x^{(0)}_\alpha P^{\prime\prime}(x^{(0)}_\alpha)=0.
\end{equation}
The polynomial
\begin{equation}
Q(x)=xP^\prime(x)-2\xi d(\xi)P^\prime(x)+\xi x P^{\prime\prime}(x),
\end{equation}
has the same degree $N$ and roots $\{x^{(0)}\}$ as $P(x)$.  Therefore it must be a multiple of $P(x)$, with the multiplication factor easily determined as $N$
\begin{equation}
xP^\prime(x)-2\xi d(\xi)P^\prime(x)+\xi x P^{\prime\prime}(x)=N P(x).
\end{equation}
A change of variables $x=-\xi y$ transforms this equation into the differential equation defining the associated Laguerre polynomials $L^k_N(y)$ of degree $k=-2d(\xi)-1$ \cite{arfken:2001}
\begin{equation}
 y\frac{d^2L^k_N(y)}{dy^2}+(k+1-y)\frac{dL^k_N(y)}{dy}+NL^k_N(y)=0.
\end{equation}
As a result, the variables $\{x^{(0)}\}$ can be related to the roots of the associated Laguerre polynomials $L^k_N(y)$ of degree $k=-2d(\xi)-1$, and the solution of the deformed RG equations in the strong-coupling limit can be written as
\begin{equation}
x_\alpha(\xi)=-g\xi y_{l(\alpha)} + \tfrac{\sum_{i=1}^m\xi d_i(\xi)2\varepsilon_i}{\xi d(\xi)}+\mathcal{O}(\tfrac{1}{g}), 
\end{equation}
with $y_{l}$ shorthand for the $l$th root of the associated Laguerre polynomial $L^k_N(y)$ of degree $k=-2d(\xi)-1$.
\bibliography{debaerdemacker_tda2rg}
\bibliographystyle{h-physrev}

\end{document}